\documentstyle[12pt,epsf]{article}
\def\lsim{\lower.5ex\hbox{$\scriptstyle\buildrel < \over \sim $}}
\def\gsim{\lower.5ex\hbox{$\scriptstyle\buildrel > \over \sim $}}

\begin{document}

\begin{center}
{\large\bf              DYNAMICS OF SAWTOOTH MAP: \\
                         1. NEW NUMERICAL RESULTS
}
 \vspace {5mm}

{\it V.V.Vecheslavov}\footnote{Email: vecheslavov@inp.nsk.su} \\[5mm]

 \vspace {5mm}

                  The G.I.Budker Institute of Nuclear Physics   \\
                          630090 Novosibirsk, Russia
\end{center}
\vspace{3mm}
        \baselineskip 0.8 cm           

{\center Abstract \\}
\hspace*{0.3cm}\parbox{14cm}
 { \small
\hspace*{5mm}

   Some results of numerical study of the canonical map with a sawtooth force
are given and discovered new unexpected dynamical effects are described. In
particular, it is shown that if the values of the system parameter $K$ belong
to the countable set determined by Ovsyannikov's theorem, separatrices of
primary resonances are not splitted and chaotic layers are not formed. One
more set of values of the parameter related to the other family of
nondestructed separatrices of primary resonances was found.
The mechanism explaining the stability of the primary
resonance separatrix in the critical regime is found and described.
First secondary resonances were studies and for them were found the
$K$-values at which their separatrices are not splitted also.
  These facts are important since the presence of the nondestructed
separatrix  of the any order resonance eliminates the possibility of a
global diffusion in a phase space. New problems and open questions occurred
in this connections whose solution can facilitate the further development
of the nonlinear Hamiltonian systems theory are presented.
}

\vspace{0.5cm}
\hspace*{1mm} PACS 05.45.+b   \\
\hspace*{7mm} {\it Keywords: } Canonical map; Separatrix of resonance; Chaotic layer;
\vspace{0.5cm}

{\bf 1.  Introduction. }

    Two-dimensional maps of the form
$$
   \overline{p}\,=\,p\,+\,K\cdot f(x)\,, \quad \overline{x}\,=\,x\,+
   \overline{p} \quad  (mod \,\, 1)                                \eqno (1)
$$
with the only parameter $K$ is long and broadly used in the nonlinear physics
as quite convenient and very informative models [1,2,3]. For example,
function $f(x)=\sin(2\pi x)$ corresponds to the so-called standard map,
to the study of which many papers are devoted and  which is used in a great
number of studies. From the very beginning, the creators of the KAM theory
noted different dynamical behavior of the systems (1) for analytical and
smooth dependences $f(x)$. The main subject here is the question of the
smoothness degree (the number of {\it l} continuous derivatives of the
force $f(x)$) at which the global chaos takes place not for any arbitrarily
small parameter of the system but only higher of some threshold value
$K > K_g$. The studies of Moser and R\H{u}ssman have shown that such
a threshold always exists at {\it l} $ > 3$ [4]. However, there is no proof
of the inverse statement that at {\it l} $ \leq 3$ there is no threshold and
the diffusion in a phase space is unlimited.

   Consider the system with the sawtooth force
$$
  f(x)=\left\{
\begin{array}{ll}
       4\,x, & \mbox{if  } x \leq 0.25\, ,\\

       4\,(0.5-x), & \mbox{if  } |0.5-x| \leq 0.25 \, ,\\

       4\,(x-1.0), & \mbox{if  } |x-1.0| \leq 0.25\, .

\end{array} \right.                                              \eqno (2)
$$
This function is asymmetric $f(-x)=-f(x)$, its period is equal to unity
and its smoothness degree is {\it l }$ =0$.

  This our work was induced by Ovsyannikov's theorem [5] that for map (1),(2)
there exist separatrices of primary resonances at exactly determined
countable set of values of the system parameter $K$ [5]. Unfortunately, this
important result is not yet published (in the later work on this
subject [6] it is not even mentioned), therefore, for the reader's
convenience we put in Appendix nearly completely text of the
communication [5].

     By the example of a pendulum we remind that the separatrix of a single
nonlinear resonance is a special trajectory separating the phase oscillations
(inside the resonance) from its rotation (out of the resonance). In fact,
these are two spatially coincided branches corresponding to the back and
forth course of time, respectively. Each branch is a continuous trajectory
with an infinite period of motion, which outcomes the position of unstable
equilibrium (saddle) and then aproaches it asymptotically.
     In the presence in the system  of other (at least one) nonlinear
resonances, the separatrix is splitted into two intersecting branches outcoming
from the saddle in opposite directions but  do not return to it.
 Free ends of the branches produce an infinite number of loops of the
infinitely increasing length which fill the narrow region along the
unperturbed separatrix forming the so-called chaotic layer [1,2,3].
The overlapping of chaotic layers of all the system resonances
just means the onset of the global chaos.

  The central point of the modern nonlinear Hamiltonian systems theory
  could be considered the
statement that the splitting of the resonance separatrix and formation on its
place of the chaotic layer in the typical (i.e. nonintegrable) Hamiltonian
system occur at almost any perturbation. It is also assumed that namely
separatrices are first destroyed since they have zeroth frequencies of motion
and the interaction of nonlinear resonances in their vicinity
are always essential [1,2,3]. As the perturbation increases, invariant curves
with irrational winding numbers vanish last (for the standard map, this
is  the "golden" number $(\sqrt{5}-1)/2$
maximally remoted  from all rational numbers [7]).

    For the system (1),(2) all looks otherwise and recently obtained
results will be given below.

{\bf 2.  Critical numbers of primary resonances.}

     The main subject of further consideration will be resonances of the
system (1),(2) and their separatrices. The periodic orbit and corresponding
resonance are adopted to denote by the ratio of integer numbers $P:Q$,
where $Q$
is the number of iterations of the map to $P$ periods of the orbit [2].
Resonances with $Q=1$ are called primary and the remaining resonances
with $Q > 1$  are called  secondary.

     One of the most unusual and estonishing features of the studied
dynamics as will be shown below is the presence of  "critical" values of the
parameter $K$ at which the resonances separatrices are not splitted and
corresponding chaotic layers are not formed. The search for such regimes is
quite a fine problem and its accurate solution requires some special
computational techniques. To this end, in the present paper we used the
earlier developed  technique of measuring the intersection angle of
separatrix branches of primary resonances  at the central homoclinic point,
which for the system (1),(2) lies always in the symmetry line  $x=0.5$
(see [8] with the description of the details and measurement results for
the standard map). The justification of such a choice is discussed below.

   \begin{figure}[t]
\epsfxsize=130mm \epsffile{ 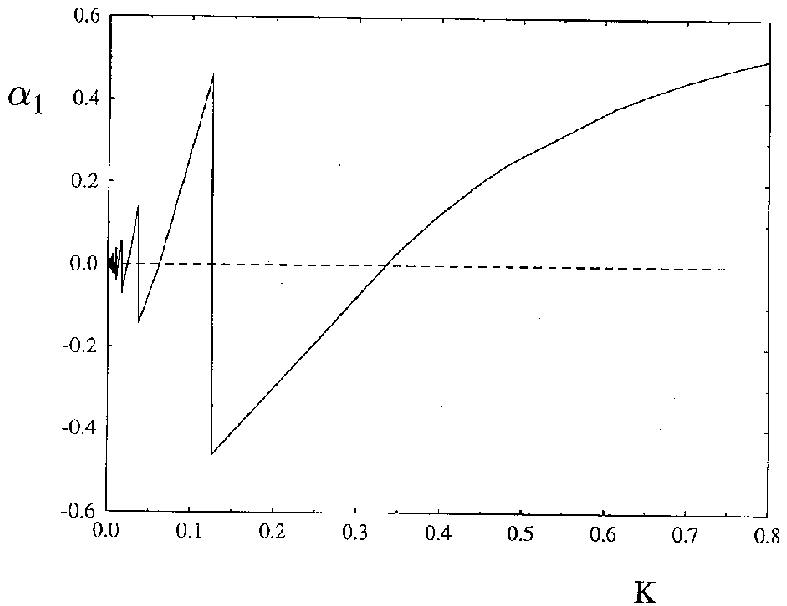}
\vspace{2mm}
\noindent
{\small Fig.1 Splitting angle $\alpha_1$ of separatrix branches of the
system (1),(2) primary resonances as a function of parameter $K$. }
  \end{figure}

     Fig.1 shows the obtained numerically for the system (1),(2) the
dependence $\alpha_1(K)$, which turned to be sign-variable and oscillating.
It is important to stress that it differs qualitatively from the well
studied by now the similar dependence for the standard map (the latter is the
sign-constant and strictly monotonical, for details see [8]). As will be
shown below, this difference is related with the essentially different
dynamical behavior of these two systems.

   Points $\alpha_1(K)=0$ in the Figure indicates the absence of splitting of
separatrices and corresponding values of the parameter $K$ we will call
critical numbers. It is seen that these numbers can be of two types: for
some of them an angle passes through zero in the process of a smooth
variation and for others in a jump. Since at the monotonical increase in
the parameter value the elements of these sets are alternate (see also Fig.3
below), we can introduce for them the through enumeration
$K_{1,m}, \,\, m=1,2,3...$  (the first index indicates the relation to the
primary resonance) and then all the odd numbers will be referred to the smooth
change of an angle and the even numbers with the jump.

   A thorough analysis showed that even critical numbers coincide exactly
with the elements of the countable set given in Ovsyannikov's theorem and
determined by the solutions of the transcendent equation (1.4) at integer
values of the coefficient $k$ (see Appendix). It turned out that found
above and not included into Ovsyannikov's theorem odd critical values are
also determined by the solutions of the same equation (1.4) but for
halfinteger values of the coefficient, therefore for any critical number we
have
$$
 K_{1,m}={\sin}^2(\beta_m/2), \,\,\, m=1,2,3,... \, ,            \eqno (3)
$$
where $\beta_m$ is the least positive root of the equation
$$
  \sqrt{2} \sin(m\beta_m/2)=\cos(\beta_m/2) \, .                  \eqno (4)
$$
The latter relations enable one, in particular, to find out exact values for
two first critical numbers $K_{1,1}=1/3$ and  $K_{1,2}=1/8$. Let us emphasize
again that these sets are referred only to primary $Q=1$ resonances of
the map (1),(2).

   \begin{figure}[t]
\epsfxsize=130mm \epsffile{ 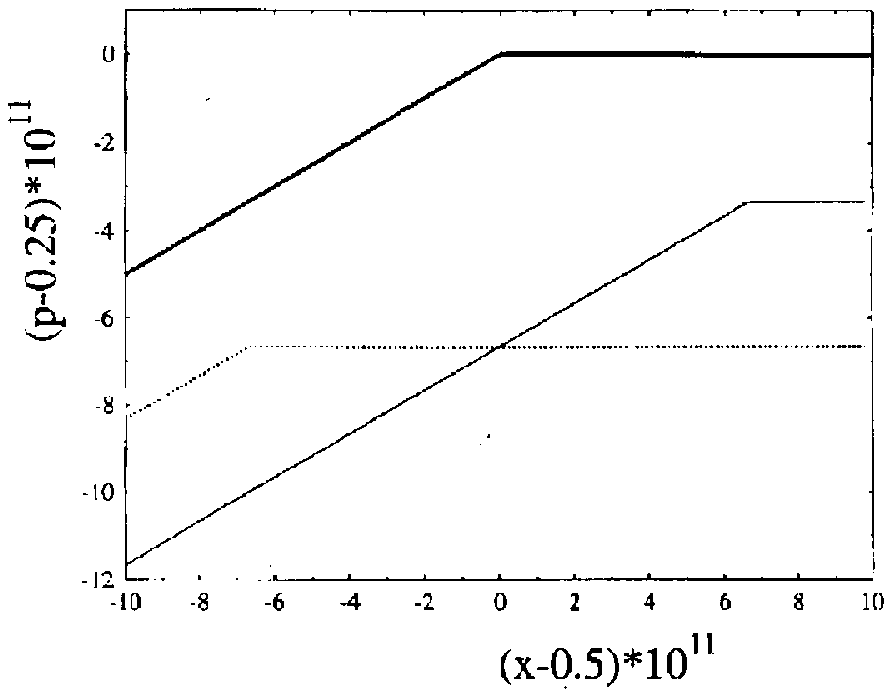}
\vspace{2mm}
\noindent
{\small
 Fig.2  The upper broken line is the section of unsplitted separatrix of the
primary resonance  at $K=K_{1,2}=0.125$. Below are the branches of the splitted
separatrix at $K=0.125-5\times 10^{-11}$.  The solid line corresponds to the
forward course of time and the dashed line to the backward course. The
branches intersection angle is $\alpha_1=0.464$.
}
  \end{figure}

     The nature of jump-like changes of the angle at even critical numbers
can be interpreted in the following way. An important part of the
Ovsyannikov theorem is the presence of exact formulae (see (1.5) in
Appendix) by which the separatrix can actually be constructed for any number
$K_{1,m}, \, m = 2,4,6,..$. All these separatrices turned to be broken lines
with one the break point coincides always with the central homoclinic point
of primary resonances in the line $x=0.5$. Fig.2 shows the picture of
intersecting branches of the primary resonance separatrix in the vicinity of
this point at the system parameter value a bit less than the critical value
$K_{1,2}=1/8$.
It is seen that the transition to the horizontal section in the forward
course of time is phase shifted to the right and in the backward course to
the left. If one constructs such a picture for $K$ a bit larger than
$K_{1,2}$, the branches exchanged  their roles and the angle conserving its
value changes its sign to the opposite sign.

    From this picture it also follows that the branch intersection angle is
quite close to the break angle value of the unsplitted separatrix (the
upper line in Fig.2) in the central homoclinic point which can be exactly
calculated. From formulae (1.5) in Appendix, it follows the angle value
asymptotical dependence on the map parameter $K$ in the
even critical points:
$$
 \alpha_{1,m} = 4\,K_{1,m}, \quad m \gg 1 \, .       \eqno (5)
$$

   \begin{figure}[t]
\epsfxsize=130mm \epsffile{ 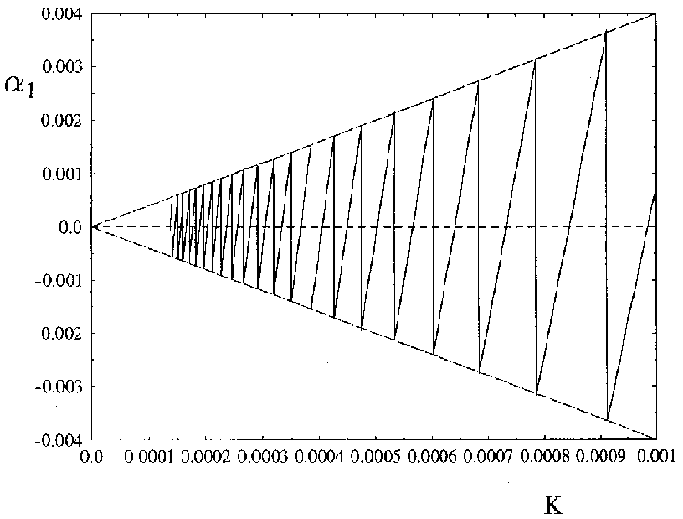}
\vspace{2mm}
\noindent
{\small
  Fig.3  Illustration of the asymptotical theoretical dependence (5)
of the break angle value of the primary resonance separatrix at the
central homoclinic point as a function of the critical number value
(see text). Inclined dashed lines are constructed by equations
$\alpha_{1} = \pm 4 K$.
}
  \end{figure}

   Fig. 3 shows the function $\alpha_1(K)$ at $K \leq 0.001 $ whose envelope
follows well the theoretical law (5). To our opinion, this fact is an evidence
of the high quality of the calculation technique we used.
As already mentioned in literature, among of all the chaos attributes
in many cases only the intersection angle of
separatrix branches can be found as accurately as required
thus facilitating the revealing of the finest details of the interaction of
nonlinear resonance and the formation of the chaotic layer [8]. The
technique can be substantially reinforced by the use of
the user specified arbitrary accuracy of calculations (in [8], for example,
the mantissa of the decimal representation of the real number had 300
significant digits).

    The found above number $K_{1,1}=1/3$ is not only
the largest of all the critical numbers. It turned out that it is at the
boundary of two regions with distinct and qualitatively different dynamical
behavior of the system (1),(2).  At $K \leq K_{1,1}$ takes place far from
trivial
dynamics whose discussion will be in the remaining part of the work. On the
contrary, at $K > K_{1,1}$ the system behavior is quite simple:  with the
growth of the parameter $K$ value, secondary resonances one by one loose
its stability
and corresponding islands of stability "submerge" in the chaotic sea. This
process is distinctly traced by the transition of eigenvalues of the
linearized motion matrix  from the complex-conjugated values (a stable
point of the elliptical type) to the real values via the equality
$\lambda_1=\lambda_2=-1$ (the hyperbolic point with reflection), which is the
evidence of the shift from the regular motion to the chaotic motion [2]. At
$K \approx 0.39$ resonances Q=3 vanish, half-primary resonances $Q=2$
vanish at $K=0.5$,
and finally, primary resonances vanish at $K=1.0$. It may turn out that for
$K>1$ the regular component of motion in the system (1),(2) is completely
absent. This property if approved is also the feature of the sawtooth
perturbation. In the standard map, for example, there exist arbitrary large
special values $K$ for which the measure of the regular component is always
higher than zero (new interesting results on the subject were recently
obtained in the work [9]).

     The established above fact of the existence of a countable set of
values of the parameter $K$ at which separatrices of the primary resonances
are not destroyed in spite of  the perturbing influence of many other
resonances
is of course of interest in itself and, to our opinion, it demands its
additional study. But of no less interest are its dynamical consequences and
we proceed to their description.

{\bf 2.  Dynamics of critical regimes.}

    Fig.4 in small window shows the region occupied by one chaotic trajectory
at the system parameter value equal to the first even critical number
$K=K_{1,2}=1/8$. The
lower and upper boundaries of this region are quite close (this definition
will be clarified) to separatrices of primary resonances
calculated by the exact Ovsyannikov's formulae. Here all contradicts to the
ideas of the "usual" dynamics.

   \begin{figure}[t]
\epsfxsize=130mm \epsffile{ 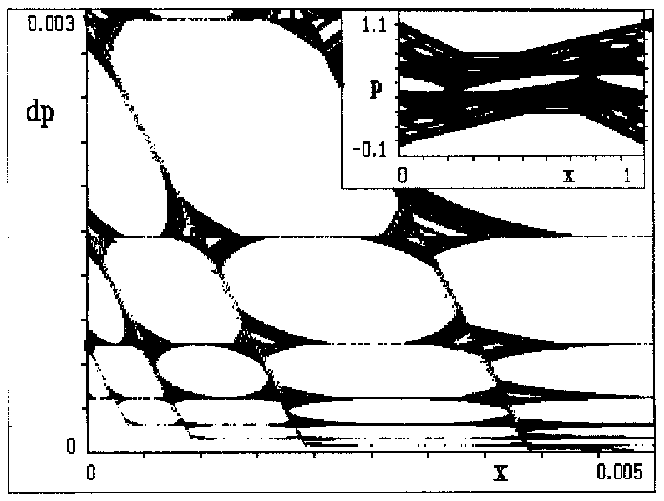}
\vspace{2mm}
\noindent
{\small
  Fig.4  The system parameter is equal to the second critical number
$K=K_{1,2} = 1/8$. Small window: the region occupied by one chaotic
trajectory with initial coordinates $x=0, p=0.37$. The number of the map
iterations is $10^{10}$.
Large window: magnified small region of the "gap" between
this trajectory and the lower separatrix of the primary resonance (see
text). Here $dp = p - p_{s}$, where $p(x)$ is the momentum of a
trajectory and  $p_{s}(x)$ is the momentum on the separatrix.
}
  \end{figure}

   It is known that there are infinitely many "untypical" Hamiltonian
systems whose separatrices of all the resonances are not splitted. These are
the so-called fully integrable systems whose dynamical behavior has no any
chaos [2,3]. A striking feature of the situation given in Fig.4 is the
coexistence of unsplitted separatrices of two adjacent primary resonances
with the region of the powerful chaos, where all the invariant curves with the
most stable irrational winding numbers are destroyed and chaotic layers
of all secondary resonances are overlapped.  We have checked also the fact
that the chaotic trajectory started inside a primary resonance remains there
during the entire counting time ($10^{10}$ iterations)
in full agreement with the Fig.3 and definition of the separatrix.

 This is apparently the main dynamical effect: at critical numbers the primary
resonances separatrices are not only destroyed  with no formation of the
chaotic layer but they form a stable invariant manyfold which does not allow
other trajectories to be intersected. Having
the full phase extension they, as "dams", divide the phase space
into sections isolated each from other. This circumstance is especially
important for applications since it prohibits the momentum global diffusion
and eliminates, for example, an unlimited rise of particle energy. As far as
we know, anything similar was never observed !

    It is relevant to note that far from critical numbers there is
apparently no unpenetrable barriers and the system behaves as "usual".
 For example, at $K = 0.3$,
the chaotic trajectory started at the initial conditions $x=0, \,\,p=0.123$
during $10^6$ iterations was occurred in the regions of seven adjacent
primary resonances.

   On the base of all mentioned above one has to recognize that the unusual
dynamics is not related to the circumstance whether the separatrix is
splitted or not but rather to that whether it is "transparent" for other
trajectories or not.
  In the recent paper [6], there is a theorem (let us call it the second
Ovsyannikov's theorem) where introduced the new and different from the
generally accepted definition of the separatrix. The new separatrix is not
splitted at any value of the system parameter $K < 1$.  But, as our studies
have shown, it is penetrable for other trajectories for almost all the
values $K$ except the critical values (3),(4). But for the critical $K$
separatrices of both kind (according to conventional viewpoint and to second
Ovsyannikov's theorem) coincide exactly with the objects of the first
Ovsyannikov's theorem [5] (see Appendix) and are unpenetrable obstacles
for other trajectories.

   The chaotic trajectory in Fig.4, as said above, approaches quite
close to the separatrices. For obtaining the quantitative estimate of this
closeness, the minimum distance in momentum $dp_{min}$ between this
trajectory and of primary resonances separatrices (calculated by exact
formulae (1.5)) was fixed. As it turned out, between the trajectory and
the lower separatrix there is a "gap" whose minimum width is
$dp_{min} \approx 3 \cdot 10^{-6}$.  A very magnified small part of the
gap is shown in large widow of Fig.4, from which it is seen that the gap is
filled in with secondary resonances of relatively high orders. The dynamics
of this region will be discussed later (see Section 5) after considering the
properties of separatrices and critical numbers of the secondary resonances.

    With the deviation of the system parameter $K$ from the critical value
$K_{1,2}$ to the side of an increase, the separatrix starts to allow other
trajectories to penetrate but the mean time $<T_{c}>$
(the number of iterations) of resonance passage depends on the detuning
value $\Delta K=K-K_{1,2} > 0$.
For finding out this dependence the following measurements were performed.
At a fixed value of the parameter $K = K_{1,2}+\Delta K $ in the region
between two adjacent resonances (see small window of Fig.4) some random
chaotic trajectories are introduced and time $T_{c}$ is fixed of their
first occurrence in the regions either below the lower or above the
upper resonances. Here it is necessary
to make the general note which is related to our work on the whole. It is
very important to choose initial conditions in such a way to have
jrajectories chaotic for sure since only these trajectories can abandon
"their" resonances and shift to others. This circumstance was checked in all
the cases by the value and  behavior of the Lyapunov exponent [2].

   \begin{figure}[t]
\epsfxsize=130mm \epsffile{ 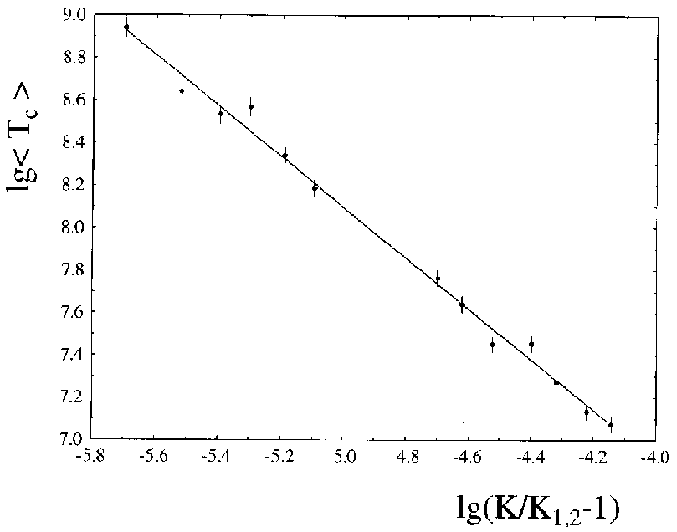}
\vspace{2mm}
\noindent
{\small
 Fig.5  Time $<T_{c}>$ required for passing the primary resonance
(mean by 100 random chaotic trajectories) as a function of the relative
deviation of the system parameter $K$ from the critical value
$K_{1,2}=1/8$ at $K > K_{1,2}$. Logarithms here are decimal.
}
  \end{figure}

   Results of these measurements are given in Fig.5 and their treatment by
the least-squares method enables us to write down an empirical formula for
the passage mean time of primary resonance:
$$
 <T_{c}> = 135 \left (\frac{K_{1,2}}{K-K_{1,2}}
  \right )^{1.193}(1 \pm 0.09),  \quad K > K_{1,2}.            \eqno (6)
$$
It is seen that with an approach to the critical regime the passage time
grows unlimitly. This circumstance enables, in principle, to use
the construction of dependencies of the kind (6) for the search for critical
numbers but such  method is much less exact and much more cumbersome than
the measurement of the separatrix splitting angle.

     It would be natural to see what is happening with the deviation of the
system parameter $K$ from the critical value $K_{1,2}$ to the side of an
decrease.  It was
found out, however,  that the dynamical situation occurred is qualitatively
different. For its understanding, it is necessary to learn the
characteristics of secondary resonances and therefore the discussions of
this dynamics will also be done in Section 5.

   \begin{figure}[t]
\epsfxsize=130mm \epsffile{ 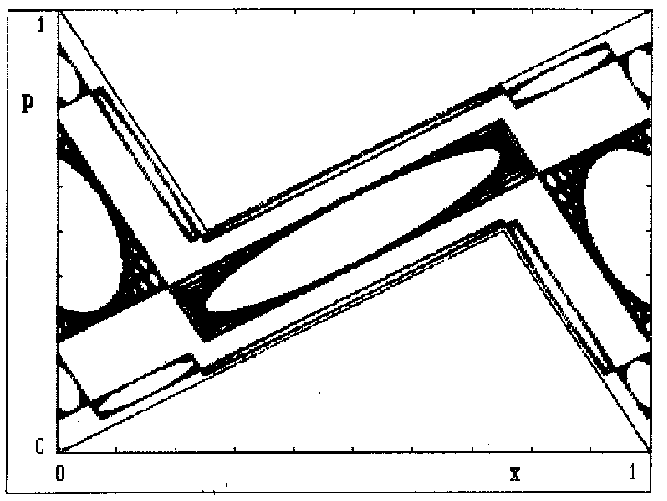}
\vspace{2mm}
\noindent
{\small
  Fig.6  The system parameter is equal to the first critical number
$K=K_{1,1}=1/3$. The lower and upper broken lines are the separatrices of
adjacent primary resonances. Three trajectories located between them with
initial coordinates providing their belonging to the chaotic components of
motion inside resonances (from below to up) 1:4, 1:2, and 3:4, respectively.
The number of the map iterations is $10^{11}$ for each trajectory.
}
  \end{figure}

    The situation with the first odd critical number looks even more
estonishing. Fig.6 constructed for the parameter value $K=K_{1,1}=1/3$ shows
separatrices of the adjacent primary resonances and three chaotic
trajectories starting inside secondary resonances 1:4, 1:2, and 3:4,
respectively. Similar calculations were carried out for resonances of other
orders but the dynamical picture always turned to be qualitatively the same:
the chaotic trajectory for the whole interval of calculation does not escape
the region of the secondary resonance inside of which it is started.  We
have shown in this Figure only three trajectories just to clarify the fact.
 The picture with many resonancs looks very attractive on the
color display screen, where each resonance is presented by its own color and
these colors are not mixed.

     Here, similarly to that given in Fig.4, all the shown region is
"closed" between separatrices of adjacent primary resonances. However, there
is an impression that each individual secondary resonance is a stable
invariant manyfold (it seems to more correct to say: the stable invariant
resonance structure) with the boundary through which the trajectories  do
not penetrate. The question is relevant "What is this boundary ?".
By the laws of the "usual" dynamics this may be a stable
invariant curve with irrational winding number, which isolates adjacent
resonances from each other [1,2,3].
But after the discussion of the situation given in Fig.4 we can rightfully
assume that this boundary might also be the unsplitted separatrix
of the secondary resonance itself. The answer to this and other questions
demands the detail analysis of secondary resonances to be done in the
nest Section.

{\bf 4.  Critical numbers of secondary resonances.}

     For a study of secondary $Q>1$ resonances we decided to use the same
techniques, which was used above for primary resonances, namely the
measurement of the splitting angle of separatrix $\alpha_{Q}$ as a
function of the system parameter $K$. Note that with an increase in $Q$
technical difficulties grow therefore, by now the reliable data are
obtained only for the first four secondary resonances $Q=2,3,4,5$.

    First of all, intersection angles of separatrix branches were measured
for the critical value of parameter $K=K_{1,2}=1/8$ and they turned to be
rather large:  $\alpha_2 \approx -0.036, \,\,\
\alpha_3 \approx -0.36, \,\,\alpha_4 \approx -0.45, \,\,
\alpha_5 \approx -1.01 $.
 This fact evidences the presence of wide chaotic layers in all  the studied
resonances, which is entirely agree with the picture shown in
small window of Fig.4.

\vspace{3mm}
\hfill       Table 1.
\begin{center}
\begin{tabular}{|c|c|c|c|c|c|}
\multicolumn{6}{c}{\bf First critical numbers of resonances.}  \\ [3mm] \hline
\multicolumn{1}{|c|}{m} &
\multicolumn{1}{|c|}{$Q = 1$} &
\multicolumn{1}{|c|}{$Q = 2$} &
\multicolumn{1}{|c|}{$Q = 3$} &
\multicolumn{1}{|c|}{$Q = 4$} &
\multicolumn{1}{|c|}{$Q = 5$} \\    \hline
 1 & $1/3                   $ & $3.3333333\cdot 10^{-1}$ & $3.3333333\cdot 10^{-1}$ & $3.3333333\cdot 10^{-1}$ & $3.3333333\cdot 10^{-1} $ \\
 2 & $1/8                   $ & $2.2949699\cdot 10^{-1}  $ & $2.9787198\cdot 10^{-1}   $ & $3.2189964\cdot 10^{-1}   $ & $3.2960547\cdot 10^{-1} $ \\
 3 & $6.1916956\cdot 10^{-2}$ & $1.7387100\cdot 10^{-1}  $ & $2.8038338\cdot 10^{-1}   $ & $3.1678073\cdot 10^{-1}   $ & $3.2801245\cdot 10^{-1} $ \\
 4 & $3.6340580\cdot 10^{-2}$ & $1.3985656\cdot 10^{-1}  $ & $2.7079692\cdot 10^{-1}   $ & $3.1415833\cdot 10^{-1}   $ & $3.2721290\cdot 10^{-1} $ \\
 5 & $2.3743290\cdot 10^{-2}$ & $1.1700662\cdot 10^{-1}  $ & $2.6505200\cdot 10^{-1}   $ & $3.1265736\cdot 10^{-1}   $ & $3.2676630\cdot 10^{-1} $ \\
 6 & $1.6679196\cdot 10^{-2}$ & $1.0061849\cdot 10^{-1}  $ & $2.6136469\cdot 10^{-1}   $ & $3.1172417\cdot 10^{-1}   $ & $3.2648587\cdot 10^{-1} $ \\
 7 & $1.2340650\cdot 10^{-2}$ & $8.8293529\cdot 10^{-2}  $ & $2.5886807\cdot 10^{-1}   $ & $3.1110653\cdot 10^{-1}   $ & $3.2630354\cdot 10^{-1} $ \\
 8 & $9.4919663\cdot 10^{-3}$ & $7.8686519\cdot 10^{-2}  $ & $2.5710423\cdot 10^{-1}   $ & $3.1067741\cdot 10^{-1}   $ & $3.2617739\cdot 10^{-1} $ \\
 9 & $7.5237127\cdot 10^{-3}$ & $7.0986237\cdot 10^{-2}  $ & $2.5581434\cdot 10^{-1}   $ & $3.1036754\cdot 10^{-1}   $ & $3.2608659\cdot 10^{-1} $ \\
10 & $6.1081453\cdot 10^{-3}$ & $6.4674926\cdot 10^{-2}  $ & $2.5350778\cdot 10^{-1}   $ & $3.1013667\cdot 10^{-1}   $ & $3.2601910\cdot 10^{-1} $ \\  \hline
\end{tabular}
\end{center}

  The next step was the calculation of critical numbers for these resonances
whose results are given in Table 1.  The data of the first column (referred
to the primary resonances) calculated by formulae (3),(4), all the
ramaining data were obtained numerically.
It turned out that each secondary resonance has its own (apparently,
countable) set of critical numbers $K_{Q,m}, \,\, m=1,2,3,...$
which, except for the first element, does not coincide with other similar
sets. And also, as in the case of primary resonances, odd critical
numbers are related to the smooth variations of the intersection angle of
separatrix branches and even numbers to the jump.

   As is seen from the Table 1, the first critical numbers $K_{Q,1}$ of the
secondary resonances coincide with the first critical number of the primary
resonance $K_{1,1}=1/3$. In order to check this estonishing fact the value
$K_{Q,1}$ was calculated for all the resonances with an accuracy of 25
true decimal digits and all of them turned to be triplets.  Actually, any
calculation results not supported by the theory can always be subjected to a
doubt because of the insufficient duration or accuracy. But the results
obtained are, to our opinion, convinceble evidences of the fact that the
unpenetrable for other trajectories boundaries of the stable structures
shown in Fig.6 are their own separatrices. Whether the equality $K_{Q,1}=1/3$
is true for any $Q$ (this would mean the unsplitting of all the separatrices)
will  be found by future studies.

    In the course of described measurements at $K_{1,1}=1/3$ it was noticed
that the winding number of the chaotic trajectory inside any resonance
during the long calculation is quite close to the value $P/Q$.  This fact can
be a base of one more method of critical numbers localization but it is
also (as the construction of dependences of the kind (6)) is less exact and
more cumbersome than the measurement of the splitting angle of separatrices.

   \begin{figure}[t]
\epsfxsize=130mm \epsffile{ 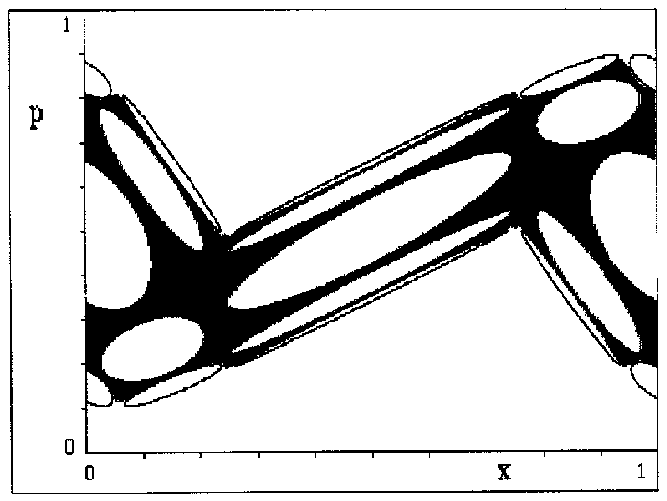}
\vspace{2mm}
\noindent
{\small
  Fig.7  The system parameter is equal to the second critical number
$K=K_{4,2}=0.321899641461... $  of secondary resonances 1:4 and 3:4 which
are presented by islands of stability. One chaotic trajectory "closed"
between unsplitted separatrices (not shown in the Figure) of these resonances.
The number of the map iterations is $10^{11}$.
}
  \end{figure}

    Fig.7 shows the "work" of critical regimes of the secondary resonances.
Here, the system parameter $K=K_{2,4}$ is equal to the second critical number of
of resonances 1:4 and 3:4 whose undestroyed separatrices "closed" the chaotic
trajectory started between them.

   It is necessary to discuss one more important feature of the system (1),
(2), which can conventionally called so: the retuning of the secondary
resonances. The map (1),(2) similarly to the standard map, has
some kind of symmetry and can be presented as product of two
involutions [2,8]. As known, this facilitates  substantially the search for
fixed points since it is necessary to examine only four lines of symmetry
enumerated for convenience in the following order: 1) $x=0;$ 2) $x=0.5;$
3) $x=p/2$ и 4) $x=(1+p)/2$ [10].

     For the standard map, the central and saddle points of any resonance in
the line of symmetry do not escape it with the change of the system parameter.
In this respect, all was turned quite different for the system (1),(2)
under study.
For example, for the resonance 1:3 at $K=K_{1,1}=1/3$ one of the
stable point is located in the second line of symmetry  $x=0.5$ whereas
in the first line $x=0$ is located the saddle.
At $K=K_{1,2}=1/8$ all turned to be just the other way around.
The dynamics of this transition is quite specific and it looks like that.
With an decrease in the parameter $K$ from 1/3, the area of the
related to the resonance 1:3 stable region decreases monotonically and
at $K=0.250$ it
vanishes completely (the value of the Green's residue calculated at
this point is equal to zero [7]).  With an additional decrease in the
parameter $K$ it occurs again and grows monotonically but the center
turned to be already in the first line and the saddle is located in
the second line.

  Positions of resonances $Q=3,4,5,6$ were traced for the first six
critical numbers from $K_{1,1}$ to  $K_{1,6}$ (see Table 1) but we still
failed to find
out here some general law. At $K_{1,1}$  one of the stable points of all
these  resonances is in the second line of symmetry. At $K_{1,2}$  the center
of the resonance 1:3 shifts to the first line and does not escape it.  At
$K_{1,2}$  the resonance 1:4 shifts to the third line and remains there. For
the resonance 1:5, the picture turned to be a little complex: at $K_{1,2}$
its center is moved to the first line, at $K_{1,3}$ it returns again to
the second line and remains there. The center of resonance 1:6 does not
escape its initial (second) line of symmetry.

    One of the most important stages in a study of the system should be the
search for and explanation of the mechanism responsible for the oscillatory
and sign-variable character of the main dependence $\alpha_1(K)$ (Fig.1)
It may turn out that the above described "movability" of secondary resonances
is related directly to the mechanism.

\newpage
{\bf 5.  Vicinity of the primary resonance separatrix at $K_{1,2}$.}

    Upon the acquaintance with the critical numbers and properties of
separatrices of secondary resonances we can return to the subjects
formulated but not considered in Section 3. Let us start with the discussion
of the dynamical situation in a "gap" between the chaotic trajectory and
separatrix of the lower primary resonance (for the brevity let us call it
the main separatrix). As is seen from large window of Fig.4, this region is
populated by resonances of relatively high orders 1:Q (the lowest still
recognizable resonance in this picture is 1:23).

     As the first step, critical numbers of some of them were found. Central
homoclinic points at odd values $Q=15,17,19...$ are in the first line of
symmetry $x=0$ that enables the use of our technique of measurement of the
separatrix intersection angle. Of the main interest are numbers closest to
the value $K_{1,2}=1/8$ for which Fig.4 is constructed. Results are given in
Table 2, where the difference between $K_{1,2}=1/8$ and critical number
$K_Q$ of the resonance $1:Q$ is indicated.
From these data it is clearly seen that with
an increase in the resonance $Q$ order, its winding number $1/Q$ and its
critical number $K_Q$ tend to the winding number (zero) and critical number
$K_{1,2}$ of the main separatrix. All these resonances have the
full phase extension and any chaotic trajectory tending to approach the main
separatrix should intersect them, which takes some certain time.

    Formula (6) from the Section 4 provides the estimate for the mean
passage time of the primary resonance, which grows infinitely with the
approach of the parameter $K$ to the critical number $K_{1,2}$ from above.
Table 2
helps the understanding the reason of this: all the secondary resonances of
the "gap" approach (from above too) to their critical numbers. If we assume
that there is some similar to the dependence (6) for these resonances, the
time required for their passage should also grow unlimitedly at
$Q \rightarrow \infty$ and $(K_{1,2}-K_Q) \rightarrow 0$.

\newpage
\vspace{2mm}
\hfill    Table 2.
\begin{center}
\begin{tabular}{|c|c|c|c|c|c|c|}
\multicolumn{7}{c}{\bf Critical numbers of the "gap" secondary resonances.
                  }  \\ [3mm] \hline
$Q          $ & $ 13              $ & $ 15              $ & $ 17               $ & $ 19               $ & $ 21               $ & $ 23               $ \\ \hline
$K_{1,2}-K_Q$ & $6.71\cdot 10^{-8}$ & $4.19\cdot 10^{-9}$ & $2.62\cdot 10^{-10}$ & $1.63\cdot 10^{-11}$ & $1.02\cdot 10^{-12}$ & $6.39\cdot 10^{-14}$ \\ \hline
\end{tabular}
\end{center}

    The direct measurement of the trajectory penetration time from the
"chaotic sea" inside the secondary resonance of high order is technically
impossible. By this reason, we find out another value, time $T_e$ required
for
the escape from the resonance region of the trajectory started inside it.
For each resonance with $Q=13, 15, 17, 19, 21$  about 500 chaotic
trajectories were calculated and the treatment of the results by the least
squares method gave the dependence
$$
   <T_{e}> = 80\, e^{0.705\, Q}(1\, \pm \, 0.07)\, ,            \eqno (7)
$$
which proves the above made assumption about infinite escape time
with $Q \rightarrow \infty$.

   These data enable us to understand the mechanism providing the stability
of the main separatrix in the critical regime $K=K_{1,2}$. The region
adjacent to the separatrix is filled with high order resonances $1:Q$ which
are located by layers the closer to the ceparatrix, the higher is the value
$Q$ (see Fig.4). Since each such resonance has the full phase extension, in
order to reach the lower layer it is necessary to intersect the upper layer.
Critical numbers of resonances with the approach to the main separatrix and
the growth of the value $Q$  approach the value $K=K_{1,2}$ (see Table 2). As a
consequence, eigen separatrices of these resonances become less and less
"transparent" for the chaotic trajectory and its escape time from them,
according to (7), grows infinitely.

    Let us turn now to the second question put in section 3 and consider
what is happening at the deviation of the system parameter $K$ from the
critical value $K=K_{1,2}$ to its decrease. From Table 2 it is clear that,
in this case, the gap resonances will be one by another successively put
into their
critical regimes and do not allow the trajectory to approach the main
separatrix at a distance (in momentum) closer than some certain $dp_{min}$.
So, at
$K=0.12499999581..$ (the critical value for the resonance 1:15)
 the minimum distance between the trajectory
and the separatrix turned to be $dp_{min} \approx 4.6 \cdot 10^{-5}$,
at $K=0.12499993294..$ (the critical value for the resonance 1:13)
$dp_{min} \approx 1.8 \cdot 10^{-4}$ with the number of map iterations
equals $10^{11}$ in every case.

   This consideration can be summarized in the following somewhat figurative
form. The higher order resonance form in front of the main separatrix some
kind of the "obstacle" resisting against an approach of the chaotic
trajectory. In the supercritical regime $K > K_{1,2}$, the trajectory manages to
intersect the primary resonance spending the more time the closer $K$
to $K_{1,2}$. At $K = K_{1,2}$ in the process of very long evolution
the trajectory can
approach the main separatrix arbitrary close but never reaches it. In
the subcritical regime $K < K_{1,2}$,  the trajectory cannot intersect the
secondary resonance located in its critical regime and approach the main
separatrix at a distance less than some certain distance.

     Whether the described scenario is typical and whether it is valid for
not only primary but for secondary separatrices is one of the open questions.
It is not yet clear also what is the onset of the supercritical and what is
the final of the subcritical regimes and where (by the parameter $K$) are
their boundaries?

{\bf 6.  Conclusion.}

   A detail study of the map (1),(2) is only started and it is too early to
draw some conclusions. But if we assume that the above described effects are
proved and generalized, the following far from simple dynamical picture takes
place.

    Each resonance from the countable set of all the system resonances is
related to its own set of critical values of the parameter $K$ at which the
intersection angle $\alpha_Q$ of separatrix branches passes through zero,
which evidences the absence of splitting and chaotic layer. The separatrix
of such a resonance (independent of its order $Q$) has the full phase
extension and is an unpenetrable barrier for other trajectories thus
eliminating the possibility of the momentum global diffusion.

    In this connection, note that in Ref.[11] in the course of the numerical
study of the map (1),(2) with the value of the parameter K=0.28625 (we draw
your attention to its closeness to the critical number $K_{3,3}$, see
Table 1) the limitation
of diffusion was noticed but not yet explained. This circumstance can
probably be understood in the course of further studies taking into account
new information.

    To our opinion, the system (1),(2) deserves its further study and we
would like to note some most important problems.

    Above mentioned facts evidence that the change of sinus by sawtooth
in the standard map, generates the qualitatively different new dynamics,
which
is beyond the frame of now existing ideas. Recall that in the
"usual" dynamics, with an increase of the parameter $K$ the transition from
the chaotic layer to the chaotic sea is studied well enough and explained by
the destruction of the invariant curves with irrational winding numbers
and by the formation of the so-called Cantorus [12]. For the system (1),
(2), this picture is apparently not valid since the main dynamical
effects turned to be related to the resonances whose winding numbers are
strictly rational. With a distance from the critical value of the parameter
the separatrix of any resonance is splitted and starts to let pass other
trajectories. Revealing of fine details of the process is apparently one of
the main and intriguing prolems.

   Another problem is related with the search for the mechanism respondible
for the oscillatory and sign-varying  character of the dependence
$\alpha_Q(K)$ (see Fig.1 for $Q = 1$)
 which mainly determines an unusual dynamics of the system.

   Of undoubted interest is also a study of the global diffusion at
 $K \leq 1/3$
taking into account of the fact that this diffusion is limited at any
critical number of any resonance. The solution of this problem seems to be
dependent of that what is the structure of the set of all the critical
numbers.

   One can hope that in the course of the search for answers to the
posed and many other questions the information can be obtained, which
enables to refine of some ideas of the modern nonlinear
Hamiltonian systems theory.

\newpage
{\bf Acknowledgements }

  The author is thankful  to L.V.Ovsyannikov for the possibility to see his
results prior to the publication and to B.V.Chirikov for numerous
discussions and advices. This work was partially supported by the Russia
Foundation for Fundamental Research, grant 97--01--00865.

\hfill    Appendix
\begin{center}
{\bf L.V.Ovsyannikov's theorem on separatrices of sawtooth map [5]
                }
\end{center}

\hspace*{7mm} Consider the equation for the function $x(t)$ defined through
the entire line
${\cal{R}}(-\infty \, < t < \, +\infty)$
$$
  x(t+h)+x(t-h)-2 x(t) = h^{2}f(x(t)) \, ,               \eqno (1.1)
$$
where $h \, > \, 0$  is the fixed constant; the function $f(x)$ is odd with
period equals 4 and it is  specified for $0 \leq x \leq 2 $  by the formula
$$
  f(x)=\left\{
\begin{array}{ll}
       \quad -x, & (0 \leq x \leq 1) \, ,\\
       x - 2, & (1 \leq x \leq 2) \, .
\end{array} \right.                                           \eqno (1.2)
$$
  The solution $x(t)$ of Eq.(1.1) with the function (1.2) is called the
separatrix if  $x(t)$ is monotonical and continuous through the entire
real axis $\cal{R}$, $x(0) = 0$ and
$x(t) \rightarrow 2$ at $t \rightarrow +\infty$,
$x(t) \rightarrow -2$ at $t \rightarrow -\infty$.

{\bf Theorem}. There exists such a sequence
$\{ h_k \}, \, k = 1,2,....$, that $h_k \rightarrow 0$ at $k \rightarrow \infty$
and for each $h = h_k$ there is the separatrix $x = x^{k}(t)$
as the solution of Eq.(1.1) with the function (1.2).
The sequence $\{ x^{k}(t) \} $ at $k \rightarrow \infty$
homogeneously on $\cal{R}$
converges to the solution -- the separatrix of the limit equation
$x^{\prime\prime}(t)=f(x)$
$$
  x(t)=\left\{
\begin{array}{ll}
       \quad \sqrt{2} \sin{t}, & (0 \leq t \leq \pi/4) \, ,\\
       2-e^{\pi/4-t}, & (\pi/4 \leq t \leq \infty) \, .
\end{array} \right.                                          \eqno (1.3)
$$
 {\bf Construction}. Let $h_k = 2 \sin(\alpha_k/2)$,
 where $\alpha_k$
 is the least positive root of the equation
$$
   \sqrt{2} \sin(k \alpha) = \cos(\alpha/2) \, .  \eqno (1.4)
$$
Then for $n h_k \leq t \leq (n+1)h_k$ the separatrix is given by formulae
$$
  x^{k}(t) = X_{n}(n+1-t/h_{k})+X_{n+1}(t/h_{k}-n) \, \,\,
  (n=0,1,2...)                                       \eqno (1.5)
$$
with constants
$$
  X_{n} = \frac{\sin(n \alpha_k)}{\sin(k \alpha_k)} \quad (n \leq k) \, ,
  \qquad
  X_{n} = 2 - {\left (\frac{\sqrt{h_k^2+4}-h_k}{2} \right )}^{2(n-k)} \quad
  (n \geq k)
$$

 Note that since the period and position of the force (1.2) in this theorem
are different from those adopted in the text (2), for the transition from
the Ovsyannikov's parameter $h$ in Eq. (1.1) to the parameter $K$ of the
map (1) and vice versa, we need to use the relation
$$
   K=h^{2}/4 \, .
$$

\begin{center}
 {\large\bf  References}
\end{center}
\vspace{3mm}

\begin{itemize}
\item[1.] B.V. Chirikov, Phys.Reports {\bf 52}, 263 (1979).
\item[2.] A. Lichtenberg and M. Lieberman, {\it Regular and Chaotic
               Dynamics}, Springer (1992).
\item[3.] G.M.Zaslavsky and R.Z.Sagdeev, {\it Introduction in nonlinear physics},
          Nauka, Moskva 1988 (in Russian).
\item[4.] J. Moser, Nachr.Akad.Wiss.,G\H{o}ttingen, Math.Phys.Kl., No 1 (1962)
     1; Stable and Random Motion in Dynamical Systems, Annals of Mathematics
     Studies No 77 (Unversity Press, Princeton 1973).
\item[5.] L.V.Ovsyannikov, private communication, May (1999).
\item[6.] L.V.Ovsyannikov, Dokl.Akad.Nauk Russia {\bf 369}, 743 (1999)
          (in Russian).
\item[7.] J. Green, J.Math.Phys., {\bf 20}  (1979) p.1183
\item[8.] V.V.Vecheslavov and B.V.Chirikov, JETP {\bf 86}, 823 (1998). \\
          V.V.Vecheslavov, JETP {\bf 89}, 182 (1999).
\item[9.]  B.V. Chirikov, Poincar\'e recurrences in microtron and the global
     critical structure, preprint Budker INP 99--7, Novosibirsk, 1999.
\item[10.] S.J.Shenker and L.P.Kadanoff, J.Stat.Phys., {\bf 4} 631 (1982).
\item[11.] B.V.Chirikov,E.Keil, and A.M.Sessler, J.Stat.Phys., {\bf 3} No 3,
           307 (1971)
\item[12.] R.S.MacKay, J.D.Meiss, I.C.Percival, Physica D, {\bf 13} 55 (1984).
\end{itemize}

\end{document}